\documentclass[lettersize,journal,onecolumn,draftclsnofoot,12pt]{IEEEtran}
\usepackage{amsmath,amsfonts}
\usepackage{algorithmic}
\usepackage{algorithm}
\usepackage{array}
\usepackage[caption=false,font=normalsize,labelfont=sf,textfont=sf]{subfig}
\usepackage{textcomp}
\usepackage{stfloats}
\usepackage{url}
\usepackage{verbatim}
\usepackage{graphicx}
\usepackage{cite}
\usepackage{amsmath}
\usepackage{multicol}
\usepackage{ragged2e} 
\usepackage{booktabs}
\usepackage{float}
\usepackage{afterpage}
\usepackage{subfig}
\DeclareUnicodeCharacter{200E}{}
\DeclareUnicodeCharacter{200B}{}
\DeclareUnicodeCharacter{200F}{}
\begin{document}

\title{Enhancing Connectivity for Emergency Vehicles Through UAV Trajectory and Resource Allocation Optimization}

\author{S.  Fatemeh Bozorgi\thanks{S. F. Bozorgi and S. M. Razavizadeh are with the School of Electrical Engineering, Iran University of Science \& Technology  (IUST), Tehran, Iran (e-mail: sf\textunderscore bozorgi‎@elec.iust.ac.ir; smrazavi@iust.ac.ir).}, S. Mohammad Razavizadeh,~\IEEEmembership{Senior Member, IEEE,} and Mohsen Rezaee\thanks{M. Rezaee is with the ICT research institute, Iran Telecommunication Research Center (ITRC), Tehran, Iran (e-mail: m.rezaeekh@itrc.ac.ir).}} 
     





\maketitle

\begin{abstract}
Effective communication for emergency vehicles—such as ambulances and fire trucks—is essential to support their operations in various traffic and environmental conditions. In this context, this paper investigates a vehicular communication system assisted by an Unmanned Aerial Vehicle (UAV), which adjusts its trajectory and resource allocation according to communication needs. The system classifies vehicles into two groups to address their varying service requirements: emergency vehicles, which require a minimum instantaneous data rate to access critical information timely, and normal vehicles. To support both categories effectively, this paper proposes a joint optimization approach that coordinates UAV trajectory planning and Dynamic Bandwidth Allocation (DBA). The objective is to maximize the minimum average data rate for normal vehicles while ensuring that emergency vehicles maintain an instantaneous rate above a predefined threshold. This approach takes into account some system constraints, including UAV propulsion power consumption, mobility limitations, and backhaul capacity. To tackle the resulting non-convex problem, an iterative optimization method is employed, where the original problem is decomposed into two subproblems: bandwidth allocation and UAV trajectory design. In each iteration, the trajectory subproblem is solved using the Successive Convex Approximation (SCA) method. Numerical results confirm that the proposed solution achieves superior performance in meeting service requirements compared to baseline methods.

\end{abstract}


\section{Introduction}
Emergency vehicles, such as ambulances, fire trucks, and police cars, must rapidly and efficiently receive critical information during crises to ensure that rescue and relief operations are carried out effectively. These vehicles require advanced communication infrastructure with high bandwidth, fast and reliable data transmission rates, and wide coverage [1]. These communcations can be as special and important case of the vehicle-to-everything (V2X) communication emerge as a crucial solution. This technology helps reduce response times, increases productivity, and decreases the likelihood of accidents in critical situations. Other benefits of V2X networks include improved road safety, accident prevention, enhanced situational awareness, reduced traffic congestion, and energy savings [2]-[4].

Vehicular communications, encompass various types of communication, including vehicle-to-vehicle (V2V), vehicle-to-infrastructure (V2I), vehicle-to-pedestrian (V2P), and vehicle-to-network (V2N). Each of these communication types plays a crucial role in enhancing road safety and transportation efficiency. For instance, V2I communication creates a link between vehicles and existing infrastructure, facilitating the transmission of essential information regarding road and traffic conditions to improve traffic management.

However, the realization of these communications depends on various factors, with one of the key factors being the efficient management of radio resources [5]. In this regard, studies such as [6-8] have focused on resource allocation in vehicular networks. [6]-[7] specifically examine the evolution and standardization of vehicular networks. 
In [8], the focus is on optimizing communication technologies in V2X systems by addressing the challenges of resource allocation through mathematical solutions. The survey also comprehensively explores various challenges and solutions related to resource management, such as fair resource allocation, interference management, and power control.

Moreover, unmanned aerial vehicles (UAVs) are increasingly utilized to enhance the capability of communication networks. The mobility and cost-effectiveness of UAVs make them highly suitable for communication in emergency conditions such as earthquakes, storms, and floods, or during times of war when traditional infrastructure is either inadequate or unavailable [9]. UAVs can be rapidly deployed and offer scalable and flexible communication solutions, which are critical in emergency scenarios. Additionally, UAVs can cater to temporary increases in users demand in specific locations [10]. Another advantage of UAVs is that as the height of the UAV increases, the likelihood of having line of sight (LOS) links with users increases [11]-[12]. This capability enhances the quality and reliability of communication networks, particularly in urban environments where buildings and other obstacles can impede signal transmission. 
In [13], UAVs are classified, and their benefits and applications are discussed. The paper also explores UAV placement, flight trajectory, backhaul data rate capacity, path planning, battery charging, spectrum allocation, and data transmission. Additionally, this survey examines constraints, optimization methods, and suitable solutions for supporting UAV-assisted wireless networks, while presenting future research directions to address the existing challenges.
In [14], a UAV network consisting of delay-sensitive and delay-tolerant users is studied. In this network, due to spectrum limitations, both backhaul and data links share the same spectrum. The main goal is to improve the data rate for delay-tolerant users while ensuring the quality of service (QoS) for delay-sensitive users. To achieve this, simultaneous optimization of bandwidth, transmission power, and the UAV's flight trajectory is performed. The UAV trajectory control, subchannel allocation, and user allocation have been studied in [15]. The objective of this research is to optimize the minimum average rate of ground users while considering data demand constraints, joint spectrum usage management, and interference control of shared channels.

Additionally, using UAVs provides several benefits that significantly enhance the performance of vehicular communication networks.  The examination of wireless networks, advanced technologies, the role of UAVs in V2X communication networks, and the challenges in this field, is covered in [16].
In [17], a UAV serves a moving vehicle, aiming to optimize the UAV's trajectory to maximize the system's energy efficiency. In this problem, an optimization issue is addressed considering the challenges arising from the vehicle's location information. In this context, a vehicle trajectory prediction framework is proposed, which includes short-term and long-term trajectory prediction models. The authors of paper [18] have investigated V2N communication in a space-air-ground network. In this paper, simultaneous optimization of transmission power and UAV trajectory is performed to maximize the data rate allocated to vehicle. Additionally, it is assumed that the vehicle moves at a constant speed, and the UAV's trajectory is determined using the successive convex approximation (SCA) method. 
In [19], UAVs are utilized as roadside units (RSUs) to facilitate data transmission to vehicles. They aim to enhance the minimum data rate of users driving on a highway by properly assigning bandwidth and also optimizing the UAV trajectory. But in [19], the UAV trajectory is obtained in one dimension and the limitations of the backhaul link and the UAV power are not taken into consideration.

Although UAV-assisted vehicular networks have attracted some attention in recent research, many existing studies overlook differences in communication requirements among vehicles. In particular, limited work has been done to address scenarios where some vehicles may require more reliable or timely data delivery due to their operational role. Motivated by this gap, we categorize vehicles into two groups: high-speed vehicles and normal-speed ones. This differentiation is especially relevant in scenarios involving emergency responders, which often travel at high speeds and require timely reception of safety-critical messages [20]. 
In addition, as vehicles move and the UAV adjusts its trajectory accordingly, variations in their relative positions influence the quality of communication links. To address this, the paper adopts a dynamic bandwidth allocation (DBA) strategy, allowing the spectrum to be adjusted in response to varying communication needs and environmental factors.
To address these challenges, we formulate an optimization problem that jointly considers UAV trajectory and bandwidth allocation under constraints such as power, UAV mobility, and backhaul capacity, aiming to improve the quality of service for both vehicle types. The main contributions of our work are outlined below:

\begin{itemize}
 \item This manuscript posits that the instantaneous data rate received by high-speed vehicles exceeds a specified threshold. This condition is crucial for the prompt and efficient delivery of safety messages, which are vital for maintaining road safety and facilitating effective traffic management.

 \item Additionally, the manuscript improves the minimum average instantaneous data rate for normal-speed vehicles. This improvement increases the stability and reliability of safety message dissemination across a range of vehicle speeds, ensuring equitable resource allocation.

 \item The study leverages DBA to flexibly distribute bandwidth among vehicles, taking into account their varying positions and service requirements.

 \item This study focuses on optimizing bandwidth allocation and UAV trajectory to maximize the minimum data rate for normal-speed vehicles. Moreover, it considers several constraints, including power limitations, UAV mobility, and backhaul link capacity, as well as the minimum instantaneous rate for high-speed vehicles. Efficient management of these factors is essential for achieving optimal resource allocation and enhancing overall operational effectiveness.
   \end{itemize}

\begin{table}[t]
\centering
\captionsetup{justification=centering, labelsep=period} 
\caption{The main notations} 
\label{tab1} 
\renewcommand{\arraystretch}{1.2} 
\label{tab1}
\begin{tabular}{| l | l |}
\hline
\textbf{Parameter} & \textbf{Definition}\\
\hline
\normalsize ${V}$ & The number of vehicles\\
\normalsize ${J}$ & The number of timeslots\\
\normalsize ${T_{D}}$ & UAV flight duration\\
\normalsize ${(x[j],y[j],z)}$ & UAV instantaneous position\\
\normalsize ${s_v}$ & ${v^{th}}$ vehicle speed\\
\normalsize ${{{\mathbf{p}}_v}[j] = [{x_v}[j],{y_v}]}$ & The instantaneous position of ${v^{th}}$ vehicle\\
\normalsize ${d_0}$ & Average channel gain at 1m\\
\normalsize ${V_1}$ & Number of high-speed vehicles\\
\normalsize ${V_2}$ & Number of normal-speed vehicles\\
\normalsize ${P}$ & Power of UAV signal transmission\\
\normalsize ${B{\kappa _{{v}}} }$ & The bandwidth allocated to the ${v^{th}}$ vehicle\\
\normalsize ${B_{BH}}$ & Backhaul link bandwidth\\
\normalsize ${P_{bs}}$ & Base station transmission power\\
\normalsize ${(x_{B},y_{B},z_{B})}$ & The base station position\\
\normalsize ${S_{U}}$ & UAV max speed\\
\normalsize ${E_x}$ & Length of the road\\
\normalsize ${E_y}$ & The width of the road\\
\normalsize ${S_{V}}$ & Maximum allowed speed\\
\normalsize ${\sigma ^2}$ & Noise power at the vehicle\\
\normalsize ${R_{th}}$ & Minimum rate for high-speed vehicles\\
\normalsize ${P_{U}}$ & Total budget for UAV power consumption\\
\normalsize ${S_{e}[j]}$ & UAV speed at $j^{th}$ timeslot\\
\normalsize ${\min_s}$ & Minimum vehicle speed\\
\normalsize ${\max_s}$ & Maximum vehicle speed\\
\normalsize ${\sigma_B^2}$ & Noise power at the UAV\\
\hline
\end{tabular}
\end{table}

Ensuring a minimum instantaneous data rate is critical for the rapid delivery of safety messages to high-speed and emergency vehicles, as it enables swift access to essential information about road hazards and traffic conditions, thereby enhancing road safety and reducing the risk of accidents. Additionally, the power consumed by UAV includes the power related to communications and propulsion, where the power required for propulsion is typically significantly greater than that required for communications [21]-[22]. Therefore, incorporating UAV power consumption into computational models is crucial for optimizing UAV trajectory planning and bandwidth allocation. Given the complexities involved, deploying UAVs in vehicular networks presents specific challenges related to user mobility and dynamic environments. This paper addresses these challenges using the SCA method and a two-stage iterative algorithm.

The structure of the paper is as follows. Section II introduces the system model and outlines the problem formulation. Section III presents the proposed algorithm. Section IV contains the numerical results that confirm the efficiency of the joint UAV trajectory and resource allocation in our scenario. Finally, the paper concludes with Section V.

\emph{Notations}: Throughout this paper, italic letters represent scalars, while bold lowercase letters signify vectors.  The 2-norm of a vector, the logarithm to the base 2, and the transpose of a vector are indicated by $||.||$, ${\log _2}(.)$, and $(.)^T$, respectively. Additionally, if $f(.)$ denotes a function, then $f(.)^L$ represents the lower bound of the function $f(.)$. A notation for partial parameter symbols is provided in Table 1.

\section{Problem Formulation}
In this paper, a UAV will be used on a road to transmit information to vehicles. Fig. 1 illustrates this system model. The present paper examines a single UAV; however, in practice, a highway can be divided into smaller sections, each of which can be covered by a UAV. Furthermore, to prevent co-channel interference, different frequencies can be considered for nearby UAVs. The total number of vehicles is indicated by ${V}$, and the total time is assumed to be ${T_{D}}$. For simplicity, we assume ${T_{D}} = J\Delta $ where $\Delta$ is considered small enough that the variation in the speed of the vehicles and the UAV can be neglected during this time slot. The number of vehicles that are in a time slot $j(j = 1,...J)$ in the determined area is ${\nu }$. The vehicle speeds are modeled as independent values drawn from a truncated Gaussian distribution over the interval ${[{\rm{ }}{\min _s}~{\rm{ }}{\max _s}]}$. Once determined, each vehicle is assumed to maintain its assigned constant speed throughout its presence in the coverage area [18]-[19],[23]-[24]. Therefore, the instantaneous position of each vehicle is calculated by ${{x_v}[j] = {x_v}[j - 1] + s_v\Delta}$, where ${s_v}$ is the speed of the ${{v^{th}}}$ vehicle, and ${{x_v}[j]}$ represents its position at the ${{j^{th}}}$ time slot along the length of the road.

\begin{figure}[]
\begin{center}
\centering
\includegraphics[trim={0cm 0.5cm 0cm 0cm},clip, width=8.5cm, height=6cm]{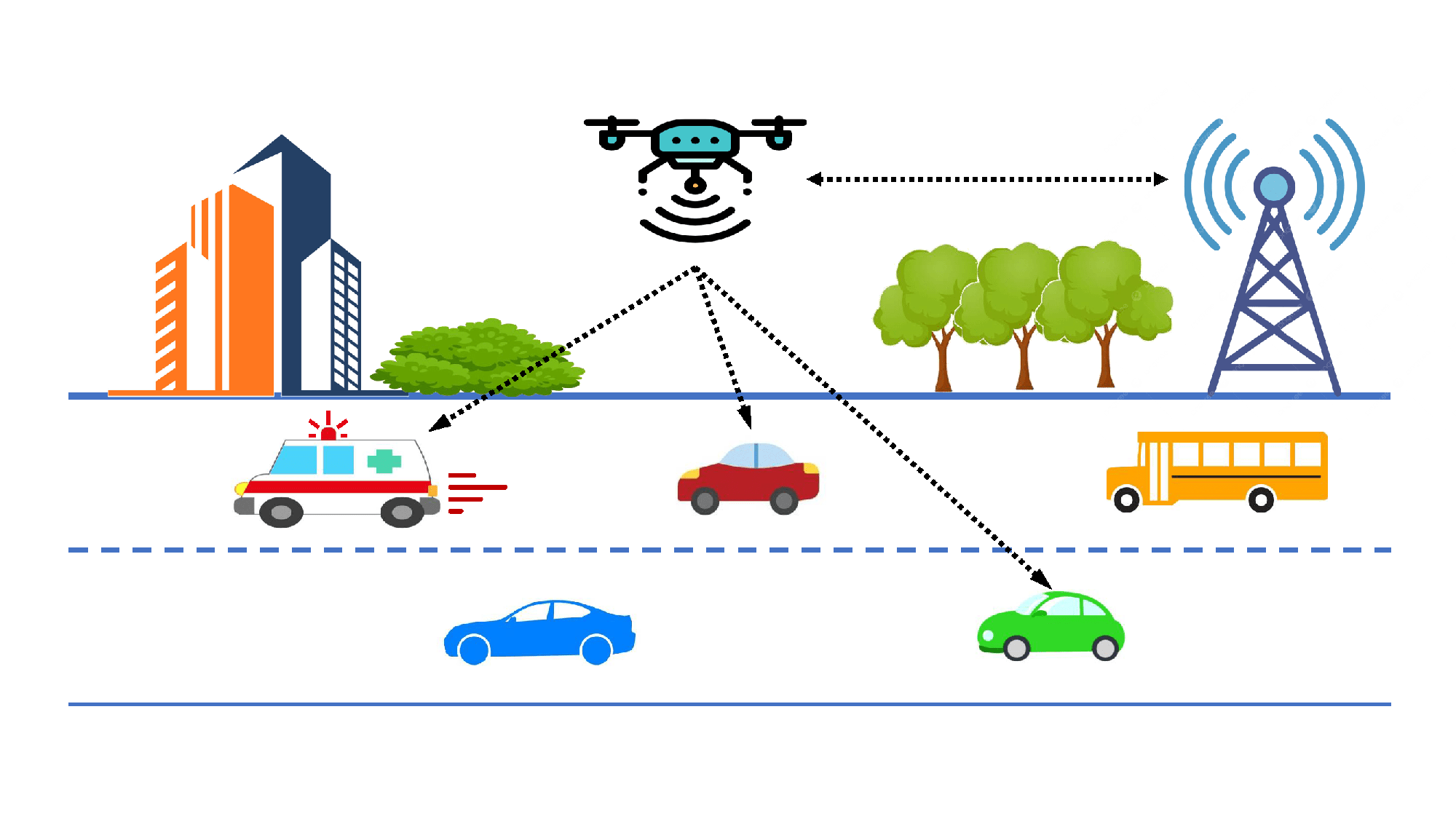}
\caption{System model. UAV trajectory in vehicular network with power and backhaul limitations.}
\label{fig1}
\end{center}
\end{figure}

The primary information about the vehicles, including location and speed, is provided to the UAV [17], [24]. Furthermore, the instantaneous position of the UAV can be represented as ${(x[j],y[j],z)}$. The altitude of the UAV is taken as a constant value ${z}$ [17]-[19],[21]-[22],[24]-[25]. Consequently, as shown in Fig. 2, the instantaneous distance between the UAV ‎and each of the vehicles is calculated as follows:

\begin{figure}[]
\begin{center}
\centering
\includegraphics[trim={0cm 0cm 0cm 0cm},clip, width=8cm, height=5.5cm]{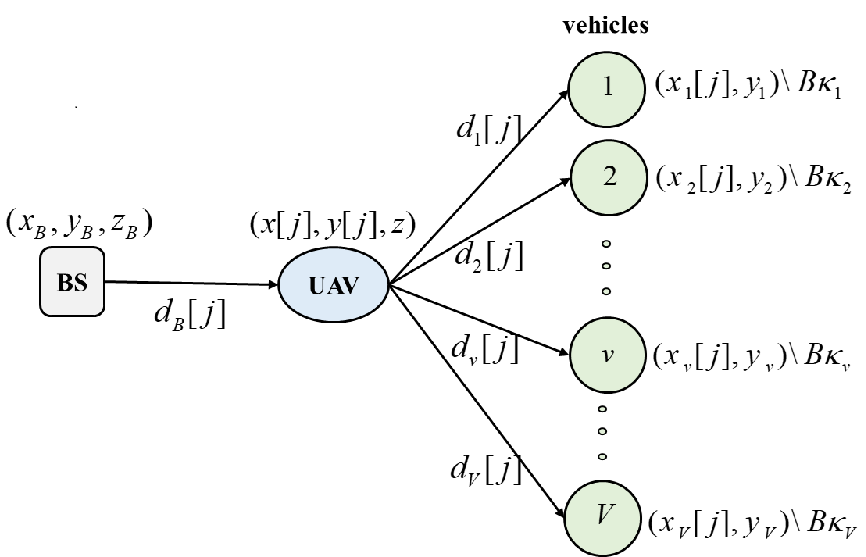}
\caption{Representation of some parameters (distance, instantaneous positions of the BS, UAV, and vehicles, and the bandwidth allocated to each vehicle).}
\label{fig1}
\end{center}
\end{figure}

\begin{equation}
\begin{array}{l}
d{_v}[j] = \sqrt {{{({x_v}[j] - x[j])}^2} + {{({y_v} - y[j])}^2} + {z^2}}={\rm}\sqrt {{\rm{||}}{{\mathbf{p}}_v}[j] - {{\mathbf{p}}_u}[j]|{|^2} + {z^2}} ,
\end{array}
\end{equation}
 where ${{{\mathbf{p}}_v}[j] = [{x_v}[j],{y_v}]}$ and ${{{\mathbf{p}}_u}[j] = [x[j],y[j]]}$ represent the position of each vehicle and the projection of UAV position on the ground, respectively at the ${{j^{th}}}$ timeslot. 

Communication between UAV and vehicles is considered LOS; ‎therefore, ‎the free-space propagation model is desirable [17]-[19],[21]-[22],[24]-[25]. Thus, the channel gain between the UAV and ‎each vehicle is as follows:‎

\begin{equation}
{h_v}[j] = d{_v}{{[j]}^{ - 2}}{d_0} ,
\end{equation}
where ${{d_0}}$ is the average gain of the channel at the reference distance ‎of 1 meter.‎

This paper assumes that ${{V_1}}$ ​vehicles have speeds greater than a certain threshold value ${S_V}$ and are classified as high-speed vehicles. Our aim is to guarantee a minimum rate for these vehicles, which are included in the set $\Phi  = \{ 1,...,{V_1}\}$. By implementing this approach, emergency vehicles can accelerate their arrival at the destination by promptly receiving essential data such as road maps and traffic information. Furthermore, the obtaining of data concerning the location of personnel on the roadway, the occurrence of accidents, and the probability of slips and falls can facilitate the enhancement of safety measures for these vehicles. The ‎objective for the other vehicles, denoted by ${v_2} \in {\hat \Phi }=\{ 1,2,...,V - {V_1}\} $, is to maximize their minimum average data rate.‎
Therefore, the instantaneous rate of high-speed vehicles and vehicles with the permitted speed, respectively, are given by:

\begin{equation}
\begin{array}{l}
{R_{v_1}[j]} = B{\kappa _{{v_1}}}[j]{\log _2}(1 + \frac{{p{d_0}}}{{{\sigma ^2}d{_{{v_1}}}{{[j]}^2}}}),{\rm{  ~~  }}\forall j{\rm{,~ }}{v_1} \in\Phi   \\
\\
{R_{v_2}[j]} = B{\kappa _{{v_2}}}[j]{\log _2}(1 + \frac{{p{d_0}}}{{{\sigma ^2}d{_{{v_2}}}{{[j]}^2}}}){\rm{,    ~~}}\forall j{\rm{, }}~{v_2} \in {\hat \Phi }
\end{array}
\end{equation}
where ${p}$ is the constant signal power transmitted from the UAV to each vehicle. The variable ${{\sigma ^2}}$ represents the power of the additive white Gaussian noise (AWGN) at the vehicle. Note that since a constant power is considered for each subcarrier, the transmitted power for each vehicle is linearly proportional to the allocated bandwidth. In addition, for simplicity, the transmitted signal power is assumed to be the same for all vehicles. The noise power is also considered equal for them. This paper assumes that the UAV employs orthogonal frequency-division multiple access (OFDMA) modulation to transmit information to vehicles and ${B{\kappa _{{v_l}}}[j],l \in \{ 1,2\} }$ is the bandwidth that is allocated to ‎the ${{{v_l}^{th}}}$‏ ‏vehicle at the ${{j^{th}}}$ time slot. In addition, ${{\kappa _{v_l}}[j]}$ is known for vehicles via the signaling channel at each time slot and should satisfy the following constraints:‎ 

\begin{equation}
\begin{array}{l}
0 \le {\kappa _{v_l}}[j] \le 1,{\rm{~~  }}\forall j,l \in \{ 1,2\}\\
\sum\limits_{v = 1}^V {{\kappa _{v}}[j]}  \le 1.{\rm{~~  }}\forall j
\end{array}
\end{equation}

Furthermore, the UAV trajectory must adhere to the following mobility constraints:

\begin{equation}
\begin{array}{l}
||{{\mathbf{p}}_u}[j] - {{\mathbf{p}}_u}[j-1]|| \le {S_{U}}\Delta ,\\
0 \le x[j] \le {{\rm{E}}_{x}},\\
0 \le y[j] \le {{\rm{E}}_{y}},
\end{array}
\end{equation}
where ${S_{U}}$ is the maximum speed of the UAV and ${{S_{U}}\Delta }$ is the maximum distance that the UAV flies during each ‎time slot.‎ It should be noted that the initial position of the UAV ($j=0$) is considered as $[x_0,y_0,z]$.‎ ${{{\rm{E}}_x}}$ indicates the end of the determined area along the length of the road and ${{{\rm{E}}_y}}$ denotes the width of the road.

Based on [25], the UAV power consumption is as follows:

\begin{equation}
\begin{array}{l}
{P_u}({S_e}[j]) = \underbrace {{P_{}}}_{{{\rm{communication}}}}+ \underbrace{{P_0}(1 + \frac{{3{S_e}{{[j]}^2}}}{{U_t^2}}) + {P_i}{(\sqrt {1 + \frac{{{S_e}{{[j]}^4}}}{{4s_0^4}}}  - \frac{{{S_e}{{[j]}^2}}}{{2s_0^2}})^{1/2}}
{+ \frac{1}{2}{d_1}\rho_1 s_rA{S_e}{[j]^3},}}_{{{\rm{power\;consumption\;of\;UAV}}}}
\end{array}
\end{equation}
where ${{S_e}[j] = \frac{{||{{\mathbf{p}}_u}[j] - {{\mathbf{p}}_u}[j - 1]||}}{\Delta }}$ indicates the speed of the UAV at different time slots. $P=pV$, ${{P_0}}$ and ${{P_i}}$ are the total transmission power of the UAV to the vehicles, the blade profile and induced power in hover, ‎respectively.‎ ${U_t^{}}$ denotes the rotor speed.‎ ${{d_1}}$ and ${\rho_1}$ represent fuselage drag ratio and the air density, ‎respectively.‎ ${s_0^{}}$, ${s_r}$ and ${A}$ are the mean rotor induced speed in hovering, the rotor ‎solidity and the rotor disc area, respectively.

The capacity of the backhaul link to the UAV in time slot $j$ is given by the following equation:

 ‎\begin{equation}
{R_{B}}[j] = {B_{BH}}{\log _2}(1 + \frac{{{P_{bs}}{d_0}}}{{{\sigma_{B}^2}d{_{B}}{{[j]}^2}}}),
\end{equation}
where ${{B_{BH}}}$, ${\sigma_B^2}$ and ${{P_{bs}}}$ are the bandwidth of the backhaul link, the power of AWGN at the UAV and the transmit ‎power of the base station, respectively. ${d{_{B}}}$ is the distance between the UAV and the base station, ‎which can be calculated as follows:‎

 ‎\begin{equation}
\begin{array}{l}
d{_{B}}[j] = \sqrt {{{({x_{B}} - x[j])}^2} + {{({y_{B}} - y[j])}^2}{+{{({z_{B}} - z)}^2}}},
\end{array} 
\end{equation}
where $(x_{B},y_{B},z_{B})$ is the position of the BS. Therefore, the optimization problem can be formulated as follows:

\begin{subequations}\label{eq:2}
\begin{align}
& \mathop{\max_{\{\kappa_{v_l}[j],\,\forall j,\,v_l,\,l \in \{1,2\}\},\,\{{\mathbf{p}}_u[j],\forall j\}}} 
\ \mathop{\min}_{v_2} \frac{1}{J} \sum_{j=1}^J B \kappa_{v_2}[j] \log_2\left(1 + \frac{p d_0}{\sigma^2 d_{v_2}[j]^2}\right) \\
& \text{s.t.} \quad B \kappa_{v_1}[j] \log_2\left(1 + \frac{p d_0}{\sigma^2 d_{v_1}[j]^2}\right) \ge R_{th}, \quad v_1 \in \Phi \\
& \sum_{v=1}^V B \kappa_v[j] \log_2\left(1 + \frac{p d_0}{\sigma^2 d_v[j]^2} \right) \le R_B[j], \quad \forall j \\
& P + P_0\left(1 + \frac{3 S_e[j]^2}{U_t^2}\right) + P_i \left(\sqrt{1 + \frac{S_e[j]^4}{4 s_0^4}} - \frac{S_e[j]^2}{2 s_0^2}\right)^{1/2} \\
& \qquad + \tfrac{1}{2} d_1 \rho_1 s_r A S_e[j]^3 \le P_U \\
& \|{\mathbf{p}}_u[j] - {\mathbf{p}}_u[j-1]\| \le S_U \Delta \\
& 0 \le x[j] \le E_x,\quad 0 \le y[j] \le E_y \\
& 0 \le \kappa_{v_l}[j] \le 1,\quad \forall j,\ l \in \{1,2\} \\
& \sum_{v=1}^V \kappa_v[j] \le 1,\quad \forall j
\end{align}
\end{subequations}

where constraint (9b) is to guarantee the minimum desired ‎rate for each high-speed vehicle ${{v_1}}$‎ and constraints (9c) and (9d) represent the limitations of the backhaul-link rate and UAV power, respectively and ${P_{U}}$ is the total budget for UAV power consumption at each time slot. The movement limitations of the UAV are outlined in constraints (9e)-(9g). The constraints related to bandwidth ‎allocation are mentioned as (9h) and (9i). Observably, the optimization problem presented above is non-convex. To address this, the variables of the problem are iteratively optimized based on the alternating optimization method.
\section{Proposed Method ‎}
To overcome the non-convexity of the initial optimization problem (9) and to facilitate its resolution, this problem is divided into two sub-problems using the block coordinate ascent (BCA) method. The initial sub-problem involves calculating the optimal bandwidth allocation for a fixed UAV trajectory. Subsequently, the second sub-problem deals with determining the UAV trajectory using the bandwidth obtained from the first sub-problem. This section examines these two sub-problems in detail.
 
\subsection{Bandwidth Optimization}
In this sub-problem, the UAV trajectory remains fixed, ‎and our objective is to determine the assigned OFDMA ‎subchannel bandwidth ${{\kappa _{v_l}}[j], l \in \{ 1,2\}}$ for each vehicle. Thus, the sub-problem is as follows:

\begin{subequations}\label{eq:2}
\begin{align}
\begin{array}{l}
\mathop {\max }\limits_{\scriptstyle\{ {\kappa _{{v_l}}}[j],\forall j,\hfill\atop
l \in \{ 1,2\} \} \hfill} \mathop {\min }\limits_{{v_2}} \frac{1}{J}\sum\limits_{j = 1}^J {B{\kappa _{{v_2}}}[j]{{\log }_2}(1 + \frac{{p{d_0}}}{{{\sigma ^2}{{d{_{{v_2}}}{{[j]}^2}}}}}){\rm{ }}} \\
~~~~~~~~~~~~~~~~~~~~~~~~~~~~~~~~~~~~~~{v_2} \in {{\hat \Phi }},j = 1,...,J{\rm{    }}\\
s.t.
\end{array}
\end{align}
 ~~~~~~~~~~~~~~~~~~~~~~~~~~~~~~~~~~~~~~~~~~~~~~~~~~~~~~ (9{\rm{b), ~(9c)}},{\rm{~ (9h) ~and~ (9i)}}{\rm{. }}
\end{subequations}

In order to facilitate the derivation of a solution to the aforementioned problem (10), we define $\eta$ as follows:

 \begin{equation}
 \begin{array}{l}
\eta  = \mathop {\min }\limits_{{v_2}} 1/J\sum\limits_{j = 1}^J {{R_{{v_2}}[j]}},
\end{array}{}
\end{equation}
and rewrite the equivalent problem as (12):
\begin{subequations}\label{eq:2}
\begin{align}
&\begin{array}{l}
\mathop {\max {\rm{ }}}\limits_{\{ {\kappa _{{v_l}}}[j],\forall j,l \in \{ 1,2\}\},\eta} \eta \\
\emph{s.t.}
\end{array}\\
&\begin{array}{l}
\frac{1}{J}\sum\limits_{j= 1}^J {{R_{{v_2}}[j]}}  \ge \eta, 
{\rm{   ~~~~~~~~   }}{v_2} \in{\hat \Phi },~j = 1,...,J
\end{array}\\
&(9{\rm{b),~ (9c)}},{\rm{ ~(9h)~ and~ (9i)}}{\rm{.}}
\end{align}
\end{subequations}

In (12), the objective function and all constraints are linear. Consequently, (12) is a convex optimization problem, which can be solved using the interior-point method [26].

\subsection{UAV Trajectory Optimization}

With fixed bandwidth allocation, the UAV trajectory can ‎be optimized by:

\begin{subequations}\label{eq:2}
\begin{align}
\begin{array}{l}
\mathop {\max {\rm{ }}}\limits_{\{ {{\mathbf{p}}_u}[j],\forall j\} ,\eta } \eta \\
\emph{s.t.}
\end{array}
\end{align}
~~~~~~~~~~~~~~~~~~~~~~~~~~~~~~~~~~~~~~~~~~~~~~~~~~~~~~~~~~~~~~~~~~~~~~~~~~~~~~ {\rm{(9b) - (9g)}}, (12{\rm{b}}){\rm{.}}
\end{subequations}

Sub-problem 2 is non-convex. Thus, to convexify the ‎problem, by introducing an auxiliary variable ${{t_{{v_2}}}}$‎, the instantaneous distance between the UAV and each ‎vehicle adhering to the speed limit is specified as follows:

 ‎\begin{equation}
d{_{{v_2}}}{[j]^2} = {({x_{{v_2}}}[j] - x[j])^2} + {({y_{{v_2}}} - y[j])^2} + {z^2} \le {t_{{v_2}}}[j].
\end{equation}

Consequently, with the auxiliary variable $t_{v_2}$, ${{R_{{v_2}}}}$ can be reformulated as follows:

 ‎\begin{equation}
{{\tilde R_{{v_2}}}[j] = B{\kappa _{{v_2}}}[j]{\log _2}(1 + \frac{{p{d_0}}}{{{\sigma ^2}{t_{{v_2}}}[j]}})}\footnote{At the optimal solution of problem (13), constraint (14) should hold with equality, otherwise $t_{v_2}$ can be reduced to increase the objective value [27].}. 
\end{equation}

 Since the function ${{\tilde R_{{v_2}}}[j]}$ with respect to $t_{v_2}$ is convex, the first-order Taylor approximation provides a lower bound [26]\footnote{For instance, the function $f(x) = \log (1 + \frac{a}{x})$ with variable $x > 0$ and constant $a \ge 0$ is convex and its first-order Taylor approximation is ${f^\prime }({x_0})(x - {x_0}) + f({x_0}) \le f(x)$, where ${f^\prime }({x_0})$ represents the derivative of function $f({x})$ at point $x_0$.}. Furthermore, the SCA method allows for an approximation of the original function  ‎${{\tilde R_{{v_2}}}[j]}$ with a simpler function at specific local points in each iteration [28]. So, during the SCA method, a lower bound  ($\tilde R_{{v_2}}^L[j]$) for ${{\tilde R_{{v_2}}}[j]}$ is obtained at the  ${{r^{th}}}$ iteration around the local point ${t_{{v_2}}^r[j]}$ as follows:

 ‎\begin{equation}
\tilde R_{{v_2}}^L[j] = B{\kappa _{{v_2}}}[j]( - \phi _{{k_2}}^{}[j]({t_{{v_2}}}[j] - t_{{v_2}}^r[j]) + \alpha _{{k_2}}^{}[j]),
\end{equation}
where‎ ${\phi _{{k_2}}^{}[j]}$ and ${\alpha _{{k_2}}^{}[j]}$ are as bellow:
 ‎\begin{equation}
 \begin{array}{l}
{\phi _{{k_2}}^{}[j] = \frac{{\log _2^e{\rm{ }}p{d_0}/{\sigma ^2}}}{{(t_{{v_2}}^r[j] + p{d_0}/{\sigma ^2})t_{{v_2}}^r[j]}}},\\[0.5 cm]
{\alpha _{{k_2}}^{}[j] = {\log _2}(1 + \frac{{p{d_0}/{\sigma ^2}}}{{t_{{v_2}}^r[j]}})}.
\end{array}{}
\end{equation}
Similarly, ${{R_{B}}}$ and ${{R_{{v_1}}}}$, respectively are converted to (18) and (19):

\begin{flalign}
&\begin{array}{l}
\tilde R_{B}^L[j] = {B_{BH}}( - {\phi _{B}}[j]({t_{B}}[j] - t_{B}^r[j])
+ {\alpha _{B}}[j]),
\end{array}\\[0.5cm]
 &\tilde R_{{v_1}}^L[j] = B{\kappa _{{v_1}}}[j]( - \phi _{k_1}^{}[j]({t_{v_1}}[j] - t_{v_1}^r[j]) + \alpha _{k_1}^{}[j]),
\end{flalign}
where the variables are defined as follows:

\begin{equation}
\begin{array}{l}
\alpha _{B}^{}[j] = {\log _2}(1 + \frac{{{P_{bs}}{d_0}/{\sigma_B^2}}}{{t_{B}^r[j]}}),\\[0.5cm]
\phi _{B}^{}[j] = \frac{{\log _2^e{\rm{ }}{P_{bs}}{d_0}/{\sigma_B^2}}}{{(t_{B}^r[j] + {P_{bs}}{d_0}/{\sigma_B^2})t_{B}^r[j]}},\\[0.5cm]
{({x_{B}} - x[j])^2} + {({y_{B}} - y[j])^2} + {({z_{B}} - z)^2} \le {t_{B}}[j],\\[0.5cm]
\alpha _{{k_1}}^{}[j] = {\log _2}(1 + \frac{{p{d_0}/{\sigma ^2}}}{{t_{v_1}^r[j]}}),\\[0.5cm]
\phi _{k_1}^{}[j] = \frac{{\log _2^e{\rm{ }}p{d_0}/{\sigma ^2}}}{{(t_{v_1}^r[j] + p{d_0}/{\sigma ^2})t_{{v_1}}^r[j]}},\\[0.5cm]
{({x_{v_1}}[j] - x[j])^2} + {({y_{{v_1}}} - y[j])^2} + {z^2} \le {t_{v_1}}[j],\\
\end{array}
\end{equation}
where ${t_{B}^r[j]}$ and ${t_{{v_1}}^r[j]}$ denote the value of ${{t_{B}}[j]}$ and ${{t_{v_1}}[j]}$, respectively at the ${{r^{th}}}$ iteration.‎

For convexification (9d), we can express the third part of this constraint using an auxiliary variable ${D[j]}$ as follows:

 ‎\begin{equation}
\begin{array}{l}
D[j] \buildrel \Delta \over = {(\sqrt {1 + \frac{{{S_e}{{[j]}^4}}}{{4s_0^4}}}  - \frac{{{S_e}^2[j]}}{{2s_0^2}})^{1/2}},{\rm{   ~~~~    }}D[j] \ge 0\\[0.5cm]
 ~~~~~~\Rightarrow {D^2}[j] = (\sqrt {1 + \frac{{{S_e}{{[j]}^4}}}{{4s_0^4}}}  - \frac{{{S_e}^2[j]}}{{2s_0^2}}),\\[0.5cm]
 ~~~~~~\Rightarrow \frac{1}{{{D^2}[j]}} = {D^2}[j] + \frac{{{S_e}^2[j]}}{{s_0^2}},
\end{array}
\end{equation}
where the last equation is derived by simple mathematical relations. Therefore, the sub-problem (13) is rewritten as follows:

\begin{subequations}\label{eq:2}
\begin{align}
&\begin{array}{l}
\mathop {\max {\rm{  }}}\limits_{\scriptstyle\{ {{\bf{p}}_u}[j],\forall j\} ,\{ D[j],\forall j\} ,\hfill\atop
{\scriptstyle\{ {t_{v_l}}[j],\forall j,l \in \{ 1,2\}\} , \hfill\atop
\scriptstyle\{ {t_{B}}[j],\forall j\},\eta \hfill}}  \eta \\
\emph{s.t.}
\end{array}\\
&\frac{1}{J}\sum\limits_{j = 1}^J {\tilde R_{{v_2}}^L[j]} \ge \eta, \\
&\tilde R_{{v_1}}^L[j] \ge {R_{th}}, \\
&\sum\limits_{j = 1}^J {\tilde R_{{v}}^L[j]}  \le {\tilde R_{B}^L[j] },  \\
&\begin{array}{l}
{P_{}} + {P_0}(1 + \frac{{3{S_e}{{[j]}^2}}}{{U_t^2}}) + {P_i}D[j] + \frac{1}{2}{d_1}\rho_1 s_rA{S_e}{[j]^3}
{\rm{   }} \le {P_{U}}, 
\end{array}\\
&{\rm{||}}{{\bf{p}}_{v_l}}[j] - {{\bf{p}}_u}[j]|{|^2} + {z^2} \le {t_{{v_l}}}[j], ~~l \in \{ 1,2\},\\
&\begin{array}{l}
{({x_{B}} - x[j])^2} + {({y_{B}} - y[j])^2} + {({z_{B}} - z)^2}
{\rm{   }} \le {t_{B}}[j],
\end{array}\\
&\frac{1}{{{D^2}[j]}} \le {D^2}[j] + \frac{{S_e^2[j]}}{{s_0^2}},\\
&D[j] \ge 0,\\
‎& \text{(9e)-(9g)}.‎
\end{align}
\end{subequations}

Constraint (22h) is obtained by relaxing the equality in (21) with inequality. It should be noted that, when the optimal solution of the problem (22) is obtained, (22h) is satisfied with equality. Otherwise, if (22h) satisfies with strict inequality for any $j$, the value of $D[j]$ can decrease to reduce power consumption until (22h) is satisfied with the equality [25].

The problem (22) is not yet convex due to the non-convex nature of the constraint (22h). The RHS of (22h) is a jointly convex function of ${S_e}[j]$ and $D[j]$, and therefore, with the first-order Taylor approximation, the lower bound of the RHS of constraint (22h) can be formulated as follows:

 ‎\begin{equation}
\begin{array}{l}
{D^2}[j] + \frac{{{S_e}^2[j]}}{{s_0^2}} \ge {D^2}^r[j] + 2{D^r}[j](D[j] - {D^r}[j])\\[0.5cm]
{\rm{~~~~~~~~~~~~~~~~~~~~~ }} + \frac{{{S_e}{{^2}^r}[j]}}{{s_0^2}} + 2\frac{{{S_e}^r[j]({S_e}[j] - {S_e}^r[j])}}{{s_0^2}}\\[0.5cm]
{\rm{ = (2}}{D^r}[j]D[j] - {D^2}^r[j]) + \frac{{{\rm{(2}}{S_e}^r[j]{S_e}[j] - {S_e}{{^2}^r}[j])}}{{s_0^2}},
\end{array}
\end{equation}
where  ${{D^r}[j]}$ and ${{S_e}^r[j]}$ are the values of the variables ${D[j]}$ and ${{S_e}[j]}$, respectively at the ${{r^{th}}}$ iteration.
Consequently, problem (22) is reformulated as (24):

\begin{subequations}\label{eq:2}
\begin{align}
&\begin{array}{l}
\mathop {\max {\rm{  }}}\limits_{\scriptstyle\{ {{\mathbf{p}}_u}[j],\forall j\} ,\{ D[j],\forall j\} ,\hfill\atop
{\scriptstyle\{ {t_{v_l}}[j],\forall j,l \in \{ 1,2\}\} , \hfill\atop
\scriptstyle\{ {t_{B}}[j],\forall j\},\eta \hfill}}  \eta \\
\emph{s.t.}
\end{array}\\
&\frac{1}{{{D^2}[j]}} \le \begin{array}{l}
{\rm{(2}}{D^r}[j]D[j] - {D^2}^r[j]) + \frac{{{\rm{(2}}{S_e}^r[j]{S_e}[j] - {S_e}{{^2}^r}[j])}}{{s_0^2}}, \\
\end{array}\\
& \text{(9e)-(9g), (22b)-(22g), (22i)}‎.
\end{align}
\end{subequations}

\begin{table}[!b] 
\centering
\caption{\textbf{Algorithm 1}}
\renewcommand{\arraystretch}{1.5} 
\begin{tabular*}{17.5pc}{@{\extracolsep{\fill}}p{16.5pc}@{}}
\toprule
Iterative algorithm for optimization Problem (13) \\
\midrule
\textbf{Initialization:} \\ \hspace*{1em}set ${r = 0}$ (${r}$ is the iteration number), \\ \hspace*{1em}initialize ${{\mathbf{p}}_u^0[j], D_{}^0[j]},$ ${t_{B}^0[j]}$, ${t_{v_{1}}^0[j], t_{v_{2}}^0[j], \forall j,v}$ \\
1: \textbf{repeat} \\ 
2: \textbf{update} $r= r + 1$. \\ 
3: With given ${{\mathbf{p}}_u^r[j], D_{}^r[j]},$ ${t_{B}^r[j]}$, ${t_{v_{1}}^r[j], t_{v_{2}}^r[j], \forall j,v}$ solve problem (24) to obtain the optimal solution as ${{\mathbf{p}}_u^{r+1}[j]}$, ${D_{}^{r+1}[j]},$ ${t_{B}^{r+1}[j]}$, ${t_{v_{1}}^{r+1}[j], t_{v_{2}}^{r+1}[j], \forall j,v}$ \\ 
4: Until the value of the objective function (13) (i.e. the mean value of ${{{R_{{v_2}}[j]}},~j = 1,...,J}$) differs from the previous iteration by less than ${\varepsilon }$. \\ 
\bottomrule
\end{tabular*}
\end{table}

  \begin{figure}[]
\centering
\begin{center}
\includegraphics[trim={0cm 0cm 0cm 0cm},clip, width=8.5cm, height=6.5cm]{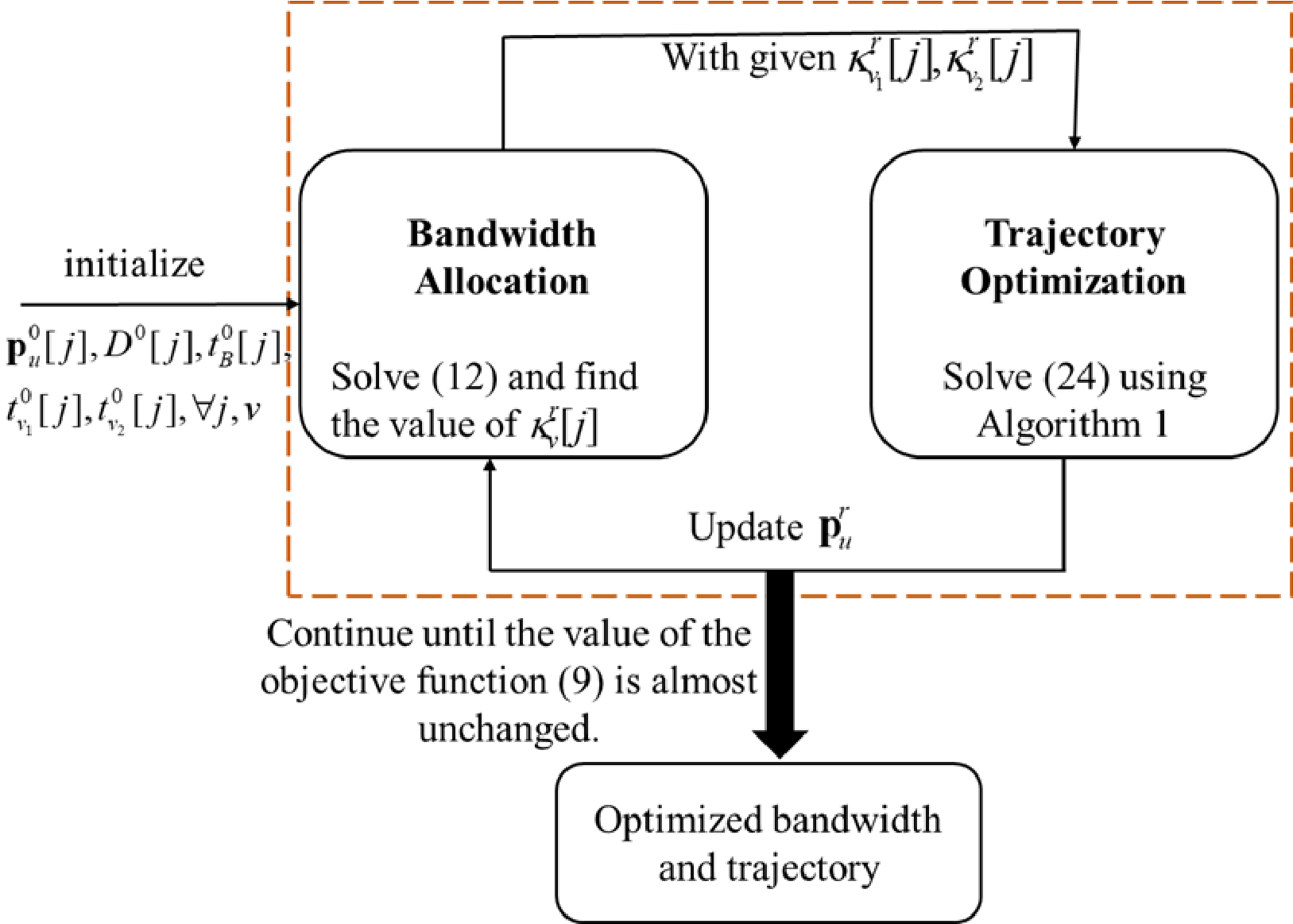}
\caption{Two-step iterative optimization method.}
\end{center}
\label{fig1}
\end{figure}

The objective function of problem (24) is linear, and all constraints are either convex or linear. Consequently, this problem is convex and can be addressed with the assistance of CVX, a convex optimization solver. Algorithm 1 describes the solution of the UAV trajectory optimization problem.

\subsection{Overall  Algorithm}
\begin{table}[]
\centering
\caption{\textbf{Algorithm 2}}
\renewcommand{\arraystretch}{1.5} 
\begin{tabular*}{17.5pc}{@{\extracolsep{\fill}}p{16.5pc}@{}}
\toprule
Proposed Algorithm for optimization Problem (9) \\
\midrule
\textbf{Initialization:} \\ \hspace*{1em}set ${r = 0}$,  \\ \hspace*{1em}initialize ${{\mathbf{p}}_u^0[j]}$, ${\kappa _{_{v_{1}}}^0[j]}$, ${\kappa _{_{v_{2}}}^0[j]}$ ${\forall j,v}$. \\
1: \textbf{repeat} \\ 
2: By solving the optimization problem (12), with the fixed trajectory, obtain ${\kappa _{_{v_{1}}}^r[j]}$, ${\kappa _{_{v_{2}}}^r[j]}$. \\ 
3: Solve the problem (24) with fixed bandwidth, using algorithm 1 to obtain ${{\mathbf{p}}_u^r[j]}$. \\ 
4: \textbf{update} $r= r + 1$. \\ 
5: Until the value of the objective function (9) differs from the previous iteration by less than ${\varepsilon }$. \\
\bottomrule
\end{tabular*}
\end{table}

To address the main problem (9) using the BCA method as shown in Fig. 3, it is necessary to decompose it into two sub-problems. Sub-problem 1, with a fixed UAV trajectory, optimizes the bandwidth of the signal sent to each vehicle. Following that, sub-problem (12) is a convex optimization problem that can be solved with the assistance of the CVX software package.

 Sub-problem 2 has been transformed into a convex optimization problem (24) with the aid of the bandwidth calculated in the preceding section and through the definition of auxiliary variables and the utilization of the SCA method. This has facilitated the optimization of the UAV trajectory. 
 The two sub-problems are solved iteratively until the increase in the objective function is less than a specified threshold. Algorithm 2 illustrates this method. ‏It should be noted that this algorithm is convergent, as detailed in the Appendix A. The worst-case complexity of this algorithm ‎with the interior-point method is ${{\rm O}({N_{it}}{(VJ)^{3.5}})}$, where ${{N_{it}}}$ represents the number of iterations required [29].

 \begin{table}[b]
\centering
\caption{UAV power consumption parameters.}
\label{tab1}
\renewcommand{\arraystretch}{1.2} 
\begin{tabular}{| c | c |}
\hline
\fontsize{9}{12}\selectfont \textbf{Parameter} & \fontsize{9}{12}\selectfont \textbf{Value} \\
\hline
\fontsize{9}{12}\selectfont Mean rotor induced velocity in hover, ${s_0^{}}$ & \fontsize{9}{12}\selectfont 5.4 m/s \\
\hline
\fontsize{9}{12}\selectfont Blade profile power, ${P_0}$ & \fontsize{9}{12}\selectfont 3.4 W \\
\hline
\fontsize{9}{12}\selectfont Rotor blade tip speed, ${U_t^{}}$ & \fontsize{9}{12}\selectfont 60 m/s \\
\hline
\fontsize{9}{12}\selectfont Induced power, ${P_i}$ & \fontsize{9}{12}\selectfont 118 W \\
\hline
\fontsize{9}{12}\selectfont Fuselage drag ratio, ${d_1}$ & \fontsize{9}{12}\selectfont 0.6 \\
\hline
\fontsize{9}{12}\selectfont Air density, ${\rho_1}$ & \fontsize{9}{12}\selectfont 1.225 $\mathrm{kg/m^3}$ \\
\hline
\fontsize{9}{12}\selectfont Rotor solidity, ${s_r}$ & \fontsize{9}{12}\selectfont 0.05 \\
\hline
\fontsize{9}{12}\selectfont Rotor disc area, ${A}$ & \fontsize{9}{12}\selectfont 0.503 $\mathrm{m^2}$ \\
\hline
\end{tabular}
\end{table}

\section{Numerical Results}
This section presents numerical results for evaluating the performance of our algorithm in trajectory and resource allocation. In the numerical results, the UAV ‎power consumption parameters are in Table 4, and Table 5 lists other parameters. Also, we set the minimum ‎rate requirement of high-speed vehicles‎ (${{R_{th}}}$) and the maximum power consumption of the UAV ‎(${{P_{U}}}$)‎, with 1000 bps and 57 dBm, respectively. There are 5 ‎vehicles in the determined area, from which 2 vehicles ‎are high-speed. In addition, the terrestrial BS is located at (-5000, 0, 30) in meters, and the UAV flight time ‎(${{T_{D}}}$) is 400 sec. Furthermore, it can be postulated that the initial position of the UAV is at point [0, 25 m, 100 m], which is similar to [19] and [24].


\begin{table}[]
\centering
\caption{Numerical results parameters.}
\label{tab1}
\renewcommand{\arraystretch}{1.2} 
\begin{tabular}{| c | c |}
\hline
\fontsize{9}{12}\selectfont \textbf{Parameter} & \fontsize{9}{12}\selectfont \textbf{Value} \\
\hline
\fontsize{9}{12}\selectfont UAV flight altitude, ${z}$ & \fontsize{9}{12}\selectfont 100 m \\
\hline
\fontsize{9}{12}\selectfont Average gain of the channel at 1 meter, ${d_0}$ & \fontsize{9}{12}\selectfont -30 dBm \\
\hline
\fontsize{9}{12}\selectfont Total bandwidth of the transmitted signal, ${B}$ & \fontsize{9}{12}\selectfont 1 MHz \\
\hline
\fontsize{9}{12}\selectfont Noise power at the vehicle, ${\sigma^2}$ & \fontsize{9}{12}\selectfont -113 dBm \\
\hline
\fontsize{9}{12}\selectfont Noise power at the UAV, ${\sigma_B^2}$ & \fontsize{9}{12}\selectfont -110 dBm \\
\hline
\fontsize{9}{12}\selectfont Transmitted signal power, ${p}$ & \fontsize{9}{12}\selectfont 0.1 W \\
\hline
\fontsize{9}{12}\selectfont UAV max speed, ${S_U}$ & \fontsize{9}{12}\selectfont 60 m/s [25], [30]\\
\hline
\fontsize{9}{12}\selectfont End of the determined area, ${E_x}$ & \fontsize{9}{12}\selectfont 10000 m \\
\hline
\fontsize{9}{12}\selectfont Width of the road, ${E_y}$ & \fontsize{9}{12}\selectfont 50 m \\
\hline
\fontsize{9}{12}\selectfont Bandwidth of the backhaul-link, ${B_{BH}}$ & \fontsize{9}{12}\selectfont 2 MHz \\
\hline
\fontsize{9}{12}\selectfont Transmission power of the base station, ${P_{bs}}$ & \fontsize{9}{12}\selectfont 46 dBm \\
\hline
\fontsize{9}{12}\selectfont Number of equal time slots, ${J}$ & \fontsize{9}{12}\selectfont 100 \\
\hline
\fontsize{9}{12}\selectfont Maximum allowed speed, ${S_V}$ & \fontsize{9}{12}\selectfont 36 m/s \\
\hline
\fontsize{9}{12}\selectfont Iteration threshold, ${\varepsilon}$ & \fontsize{9}{12}\selectfont $10^{-4}$ \\
\hline
\fontsize{9}{12}\selectfont Minimum vehicle speed, ${\text{min}_s}$ & \fontsize{9}{12}\selectfont 22 m/s \\
\hline
\fontsize{9}{12}\selectfont Maximum vehicle speed, ${\text{max}_s}$ & \fontsize{9}{12}\selectfont 40 m/s \\
\hline
\end{tabular}
\end{table}

\begin{figure*}[]
  \centering
  \begin{minipage}[b]{0.45\textwidth}
    \centering
    \includegraphics[trim={3.4cm 0.1cm 20.5cm 0.7cm},clip,width=\linewidth, height=8cm]{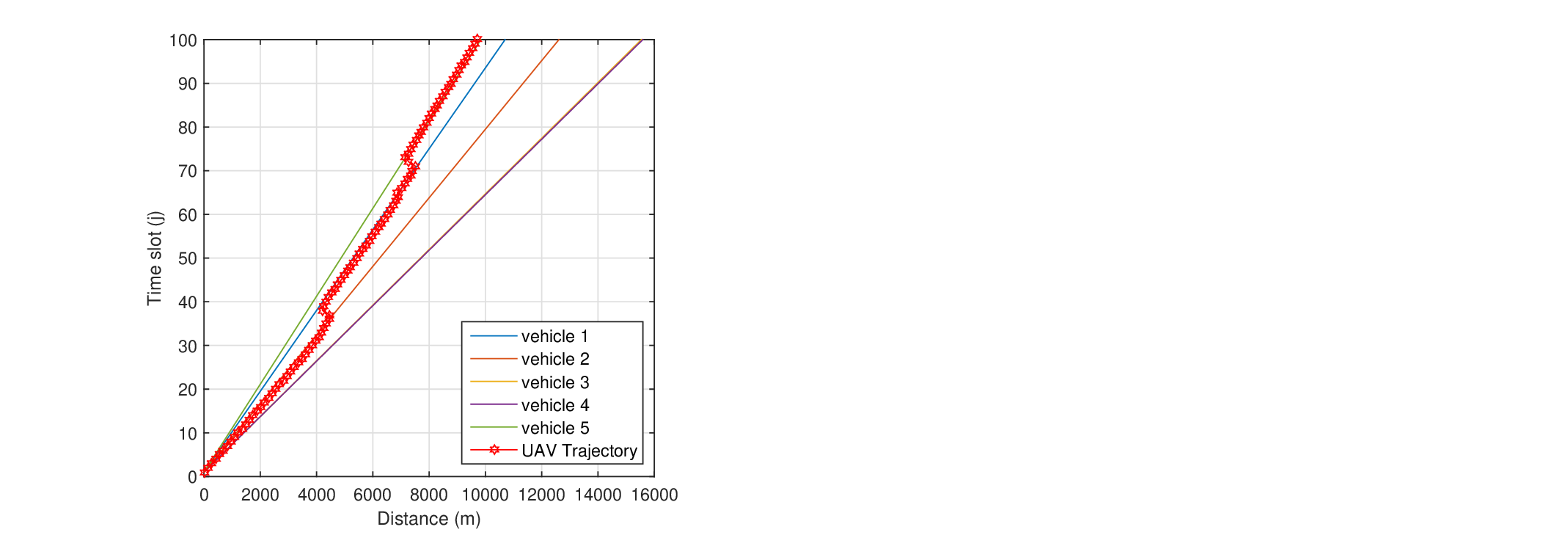}
    \vspace{0.5em}

    (a) 1D trajectory
  \end{minipage}\quad
  \begin{minipage}[b]{0.45\textwidth}
    \centering
    \includegraphics[trim={3.5cm 1cm 19.4cm 0.5cm},clip,width=\linewidth, height=8cm]{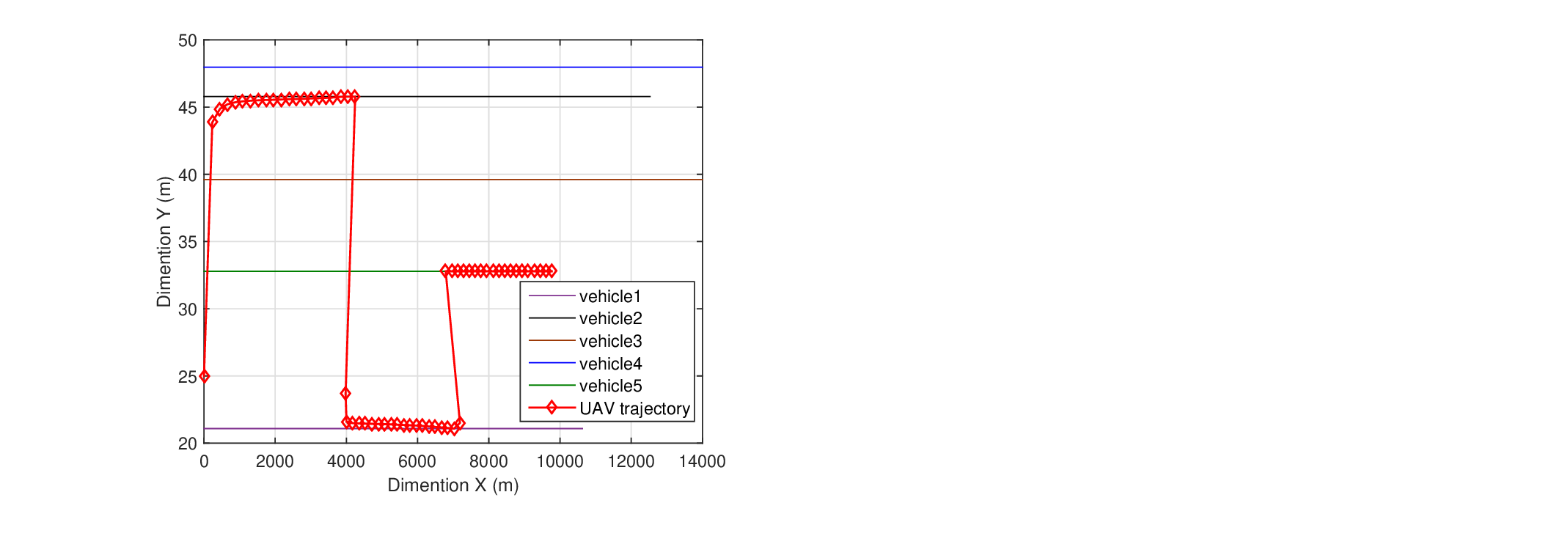}
    \vspace{0.5em}

    (b) 2D trajectory
  \end{minipage}
  \caption{Optimized UAV trajectory: (a) 1D, (b) 2D trajectory.}
  \label{fig1}
\end{figure*}

 \begin{figure}[]
\centering
\includegraphics[trim={3.2cm 0cm 19.5cm 0.4cm},clip, width=8.5cm, height=8cm]{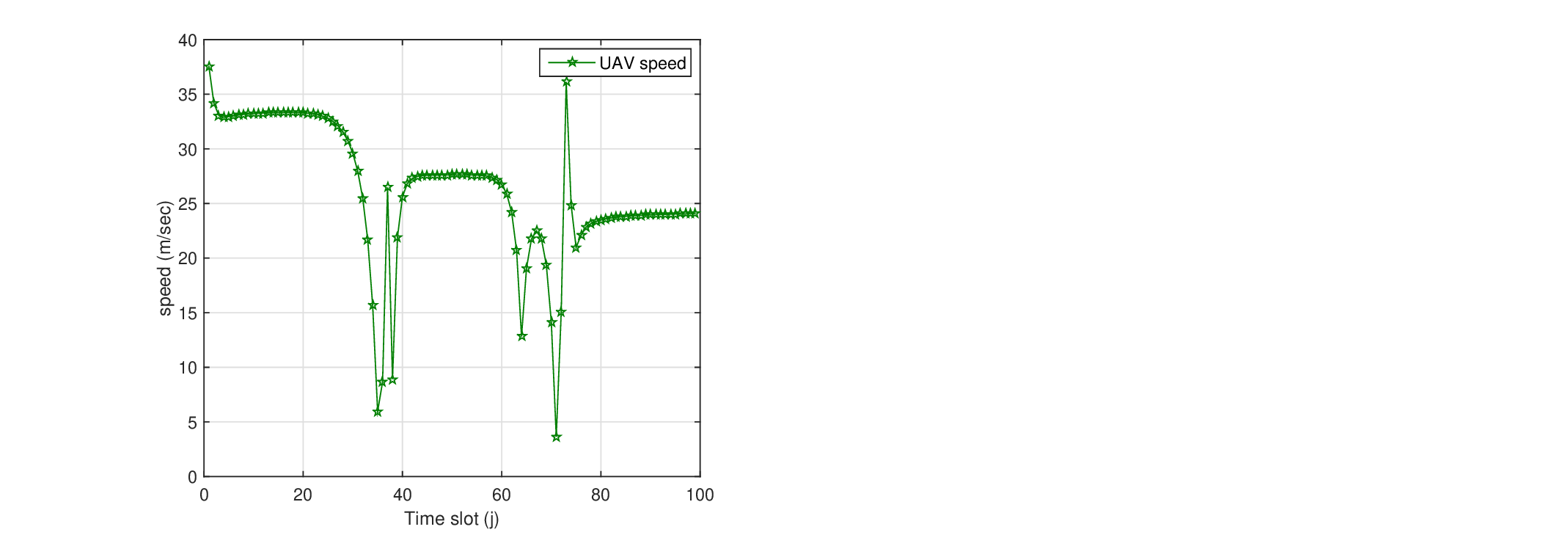}
\caption{UAV speed.}
\label{fig1}
\end{figure}

\begin{figure}[]
\centering
\includegraphics[trim={3.2cm 0.75cm 18.5cm 0.4cm},clip, width=8.5cm, height=8cm]{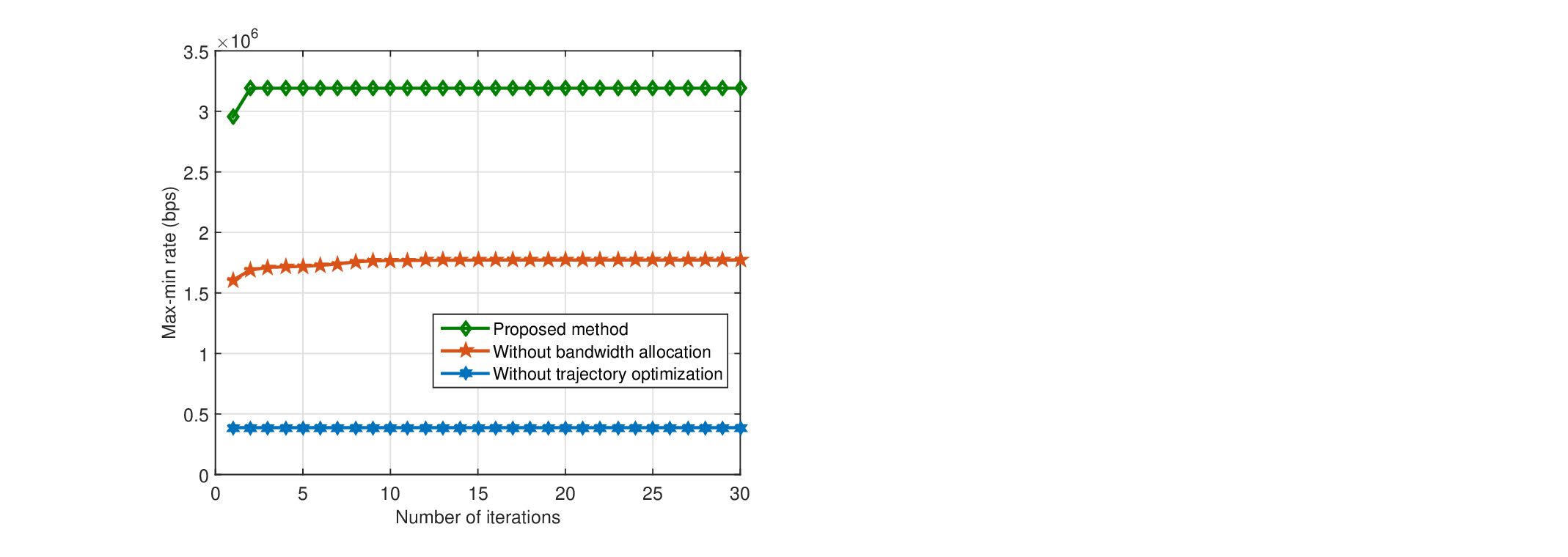}
\caption{Convergence of the proposed algorithm.}
\label{fig1}
\end{figure}

Fig. 4a demonstrates that the UAV determines its trajectory according to the QoS parameters of each vehicle.
As mentioned, the aim is to guarantee that the data rate for high-speed vehicles surpasses a specified threshold and to optimize the minimum average rate for other vehicles. In this regard, concerning the specified time ${{T_{D}}}$, the UAV approaches the vehicles that are moving at normal speeds. Initially, the UAV approaches the vehicle that has the fastest speed in comparison to the other two vehicles, and then it approaches the vehicle that has the lower speed respectively. The approach of the UAV to the vehicles will enhance the condition of the channel and facilitate communication. Therefore, according to the objective of the problem, the minimum average rate of normal-speed vehicles is maximized. Fig. 4b illustrates the trajectory of the UAV and the vehicles at specified time (${{T_{D}}}$) in two dimensions. It is observed that during this time, the high-speed vehicles (after receiving their service) have exited the determined area (10,000 meters) and are at a significant distance from it. The UAV also follows other vehicles according to the maximum-minimum problem. This approach allows for the servicing of all vehicles during the specified time (${{T_{D}}}$) and before they exit the determined area.  
  
 Fig. 5 depicts the speed of the UAV in different time slots. As demonstrated in this illustration, the speed of the UAV is not constant throughout the flight time and varies according to the vehicles in different time slots.

   Fig. 6 presents the minimum average rate of vehicles with the normal speed in three different modes for comparative analysis:
\begin{enumerate}
 \item Proposed method: joint trajectory optimization and bandwidth allocation (by algorithm 2);
  \item Without trajectory optimization: In this instance, the UAV is situated in the center of the road (${x[j]} = 5000$ m) and only bandwidth allocation has been carried out, based on the assumption in [31];
 \item Without bandwidth allocation: In this case, the trajectory of the UAV is optimized, while the bandwidth is distributed equally among the vehicles (${k_v} = 1/V$);
\end{enumerate}

As illustrated in this figure, with an increase in the iterations of the overall algorithm (2), the minimum average rate for normal-speed vehicles improves and gradually converges. In scenario 2, the SCA method is executed, while in scenario 3, since only bandwidth allocation is performed, the optimal response is a constant value. Additionally, the proposed algorithm demonstrates a significant improvement in the minimum rate compared to the two scenarios of optimizing only the trajectory (2nd curve) and only the bandwidth (3rd curve).


 \begin{figure}[]
\centering
\includegraphics[trim={3.2cm 1.75cm 20.5cm 0.4cm},clip, width=8.5cm, height=8cm]{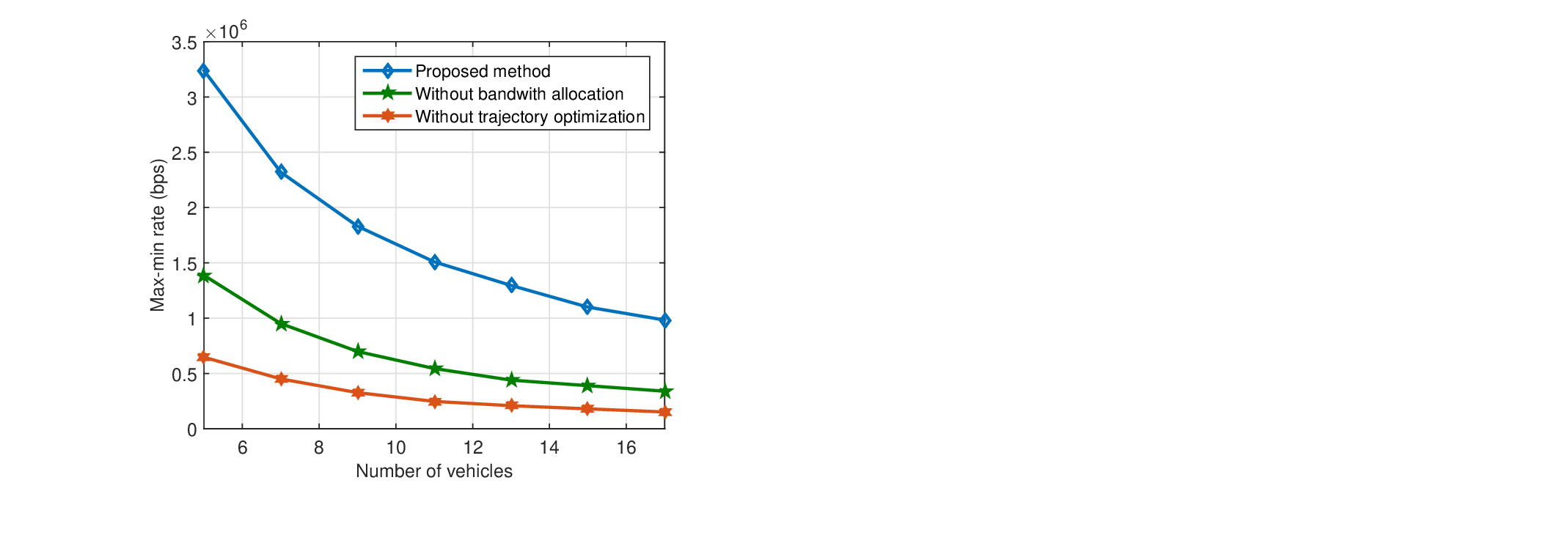}
\caption{Minimum average rate of vehicles with normal speed versus number of vehicles.}
\label{fig1}
\end{figure}

  \begin{figure}[]
\centering
\includegraphics[trim={2.8cm 0.7cm 18.5cm 0.4cm},clip, width=8.5cm, height=8cm]{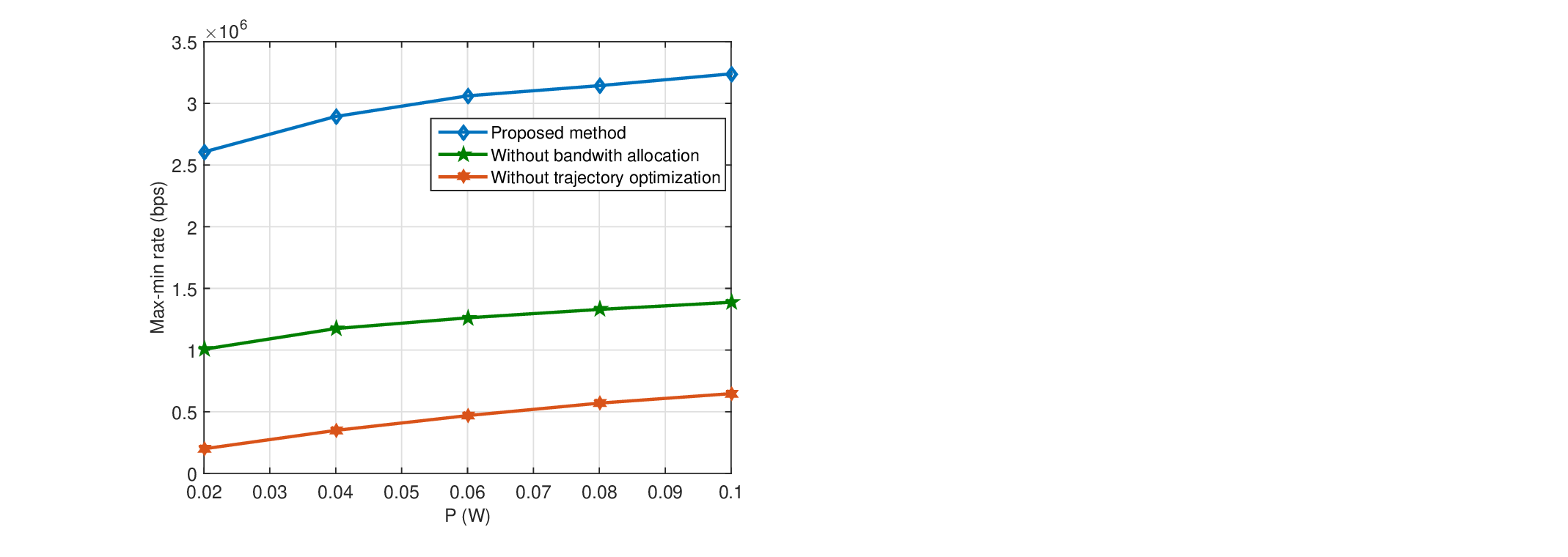}
\caption{Minimum average rate of vehicles with normal speed versus different transmit power.}
\label{fig1}
\end{figure}

In Fig. 7, the number of vehicles is from 5 to 17, it is assumed that one vehicle at the normal speed and one high-speed vehicle are added to the number of vehicles. It should be noted that averaging is done using the Monte Carlo method. As observed, with proper bandwidth allocation, the minimum average rate has been significantly increased (approximately 2 times). Moreover, our observation reveals that the value of the objective function in the case of no bandwidth allocation is greater than no trajectory optimization. Consequently, it can be concluded that trajectory optimization alone yields greater benefits than solely bandwidth allocation. Furthermore, as the number of vehicles increases, the average rate decreases. This is because available resources, such as bandwidth and UAV mobility, need to be distributed among a greater number of vehicles. Consequently, the average rate for each vehicle decreases.

The minimum average rate versus UAV transmission power for three modes, the proposed method, without bandwidth allocation mode and without trajectory optimization is shown in Fig. 8. As can be observed, the proposed method demonstrates superior performance in comparison to the other two models, with a higher minimum average rate. This indicates the significance of joint optimization of the trajectory and bandwidth. For example, if the transmission power of a UAV is 0.06 W, in the absence of bandwidth allocation, the minimum average rate is reduced by approximately 2 times, and if the scenario without trajectory optimization is employed, the minimum rate is reduced by approximately 6 times. Also, by increasing the transmitted signal power, the minimum average rate increases in all three scenarios.

 Fig. 9 illustrates the impact of varying ${{R_{th}}}$ (the minimum instantaneous rate that high-speed vehicles must receive) levels on the minimum average rate of vehicles adhering to the speed limit. In this scenario, it is assumed that one of the two high-speed vehicles is moving at a speed greater than 38 m/s. Additionally, the other vehicle moves at a high-speed, although not exceeding 38 m/s. This vehicle has been assigned an ${{R_{th_1}}=10^3}$bps. The results demonstrate that as ${{R_{th}}}$ increases, the minimum average rate declines. This is because as ${{R_{th}}}$ rises, a greater bandwidth should be allocated to high-speed vehicles, and the UAV is subject to more restrictions in approaching other vehicles, as discussed in Appendix B. Consequently, by limiting the scope for resource allocation, an increase in ${{R_{th}}}$ will result in a reduction in the average rate of vehicles that are moving at the permitted speed.

 \begin{figure}[]
\centering
\includegraphics[trim={3cm 1.2cm 19.5cm 0.4cm},clip, width=8.5cm, height=8cm]{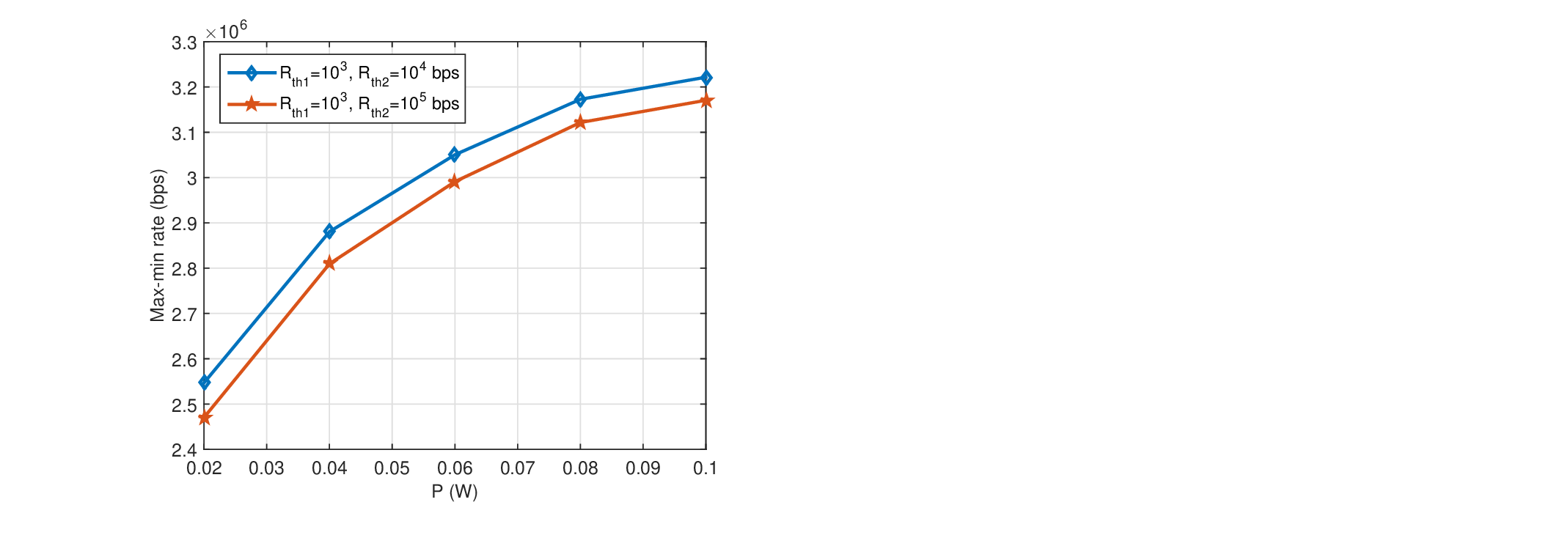}
\caption{Minimum average rate of vehicles at normal speed for different ${{R_{th}}}$​ values.}
\label{fig1}
\end{figure}
 
\section{Conclusion}

This paper focused on optimizing UAV trajectory and bandwidth allocation to each vehicle, taking into account constraints such as power, UAV mobility, and backhaul-link rate limitations. The purpose of this paper was to provide a minimum instantaneous rate for emergency vehicles that are typically driven at high speeds. Also, for other vehicles, the minimum average rate was maximized. Due to the non-convex properties of this problem, we proposed an algorithm that decomposes it into two sub-problems, and an interior-point method is used to solve each of them. Finally, by solving these two sub-problems iteratively, the sub-optimal solution to the main problem was determined. Numerical results indicate that optimizing the trajectory and effectively allocating the bandwidth significantly improves the objective of the problem. Notably, the proposed framework can be readily adapted to a variety of UAV-assisted vehicular networking scenarios characterized by different application requirements or deployment settings.

\begin{center}
\textbf{Acknowledgment}
\end{center}

The authors would like to thank the ICT Research Institute (Iran
Telecommunication Research Center – ITRC) for their valuable support
in this research.

\section{Appendix A}

The objective function for problem (9) is formulated as $f({\mathbf{k}}_{{v_2}}^r,{\mathbf{Z}}_{}^r)$ during the $r^{th}$ iteration, where the vector ${\mathbf{k}}_{{v_2}}^r$ and the matrix ${\mathbf{Z}}_{}^r$ are defined as follows:

\begin{equation}
\begin{array}{l}
{\mathbf{k}}_{{v_2}}^r = [\kappa _{{v_2}}^r[1],...,\kappa _{{v_2}}^r[j],...,\kappa _{{v_2}}^r[J]],\\[0.5 cm]
{\mathbf{Z}}_{}^r = [{\mathbf{p}}_u^r{[1]^T},...,{\mathbf{p}}_u^r{[j]^T},...,{\mathbf{p}}_u^r{[J]^T}].
\end{array}
\end{equation}

 At the local optimal value given by ${\mathbf{k }}_{{v_2}}^{r+1}$, the following relation holds in step 2 of algorithm 2:

\begin{equation}
f({\mathbf{k}}_{{v_2}}^r,{\mathbf{Z}}_{}^r) \le f({\mathbf{k}}_{{v_2}}^{r + 1},{\mathbf{Z}}_{}^r).
\end{equation}



Moreover, based on the definitions of ${\mathbf{x}}_{{v_2}}^r = [t_{{v_2}}^r[1],...,t_{{v_2}}^r[j],...,t_{{v_2}}^r[J]]$ and the function $\tilde f({\mathbf{k}}_{{v_2}}^{r},{\mathbf{x}}_{{v_2}}^r)$, which pertains to function (15), the following relationships are derived by step 3 of algorithm 2:

\begin{equation}
\begin{array}{l}
f({\mathbf{k}}_{{v_2}}^{r + 1},{\mathbf{Z}}_{}^r)\mathop  = \limits^a \tilde f({\mathbf{k}}_{{v_2}}^{r + 1},{\mathbf{x}}_{{v_2}}^r)\\[0.5 cm]
\mathop  = \limits^b {{\tilde f}^L}({\mathbf{k}}_{{v_2}}^{r + 1},{\mathbf{x}}_{{v_2}}^r)\mathop  \le \limits^c {{\tilde f}^L}({\mathbf{k}}_{{v_2}}^{r + 1},{\mathbf{x}}_{{v_2}}^{r + 1})\\[0.5 cm]
\mathop  \le \limits^d \tilde f({\mathbf{k}}_{{v_2}}^{r + 1},{\mathbf{x}}_{{v_2}}^{r + 1}) = f({\mathbf{k}}_{{v_2}}^{r + 1},{\mathbf{Z}}_{}^{r + 1})
\end{array}
\end{equation}
where (a) is justified because the optimal solution of problem (13) concerning ${\mathbf{Z}}_{}^r$  is equal to the optimal solution of it with ${\mathbf{x}}_{{v_2}}^r$[27]. 
Meanwhile, (b) follows from the tightness of the Taylor approximation (15) at the local point $({\mathbf{k}}_{{v_2}}^{r + 1},{\mathbf{x}}_{{v_2}}^r)$ [32]. Additionally, (c) is valid as a problem (24) is optimally solved with the solution ${\mathbf{x}}_{{v_2}}^{r + 1}$. Finally, (d) arises from the fact that  problem (24) gives a lower bound solution to the original problem (13). Therefore following analysis, we can conclude that:
\begin{equation}
f({\mathbf{k}}_{{v_2}}^r,{\mathbf{Z}}_{}^r) \le f({\mathbf{k}}_{{v_2}}^{r + 1},{\mathbf{Z}}_{}^{r + 1})
\end{equation}

This illustrates that, after each iteration of our algorithm, the objective function in problem (9) exhibits a monotonically non-decreasing behavior, thereby ensuring the convergence of the algorithm.

\section*{Appendix B}

In this appendix, we analytically examine the effect of the minimum rate requirement $R_{th}$ on the UAV trajectory and bandwidth allocation strategies.

Taking the derivative of $R_{v_1}[j]$ in (3) with respect to distance yields:

\begin{align}
\frac{\partial R_{v_1}[j]}{\partial d_{v_1}[j]} 
&= B \cdot \kappa_{v_1}[j] \cdot \frac{d}{d d_{v_1}[j]} \log_2\left(1 + \frac{p d_0}{\sigma^2 d_{v_1}^2[j]} \right) \nonumber \\
&= - \frac{2 B \cdot \kappa_{v_1}[j] \cdot p d_0}{\ln 2 \cdot \sigma^2 \cdot d_{v_1}^3[j] \cdot \left(1 + \frac{p d_0}{\sigma^2 d_{v_1}^2[j]} \right)} < 0.
\label{eq:rate_derivative}
\end{align}

This confirms that the achievable rate is a strictly decreasing function of distance. Hence, higher $R_{th}$ values require a reduction in $d_{v_1}[j]$, thereby tightening the UAV's trajectory constraints.

Additionally, by isolating $\kappa_{v_1}[j]$ in the rate expression, we obtain:
\begin{equation}
\kappa_{v_1}[j] \ge \frac{R_{th}}{B \cdot \log_2\left(1 + \frac{p d_0}{\sigma^2 d_{v_1}^2[j]} \right)}.
\label{eq:kappa_lower_bound}
\end{equation}

This expression indicates that increasing $R_{th}$ necessitates a corresponding increase in the allocated bandwidth $\kappa_{v_1}[j]$, unless the UAV moves closer to the target vehicle.

Since the total available bandwidth is limited, allocating more bandwidth to high-speed vehicles to meet the $R_{th}$ threshold reduces the bandwidth available to other users, particularly normal-speed vehicles.

These observations suggest that increasing $R_{th}$ can affect the optimization in two possible ways: 
\begin{itemize}
    \item It may force the UAV to reduce its distance to high-speed vehicles, thereby tightening the trajectory feasibility region.
    \item It may require a larger share of bandwidth to be allocated to high-speed vehicles, reducing the available bandwidth for other vehicles.
\end{itemize}


\begin{thebibliography}{1}

\bibitem{ams}
M. U. Ghazi, M. A. Khan Khattak, B. Shabir, A. W. Malik and M. Sher Ramzan, "Emergency Message Dissemination in Vehicular Networks: A Review," \emph{IEEE Access}, vol. 8, pp. 38606-38621, Feb. 2020.

\bibitem{ams}
C. R. Storck and F. Duarte-Figueiredo, "A Survey of 5G Technology Evolution, Standards, and Infrastructure Associated With Vehicle-to-Everything Communications by Internet of Vehicles," \emph{IEEE Access}, vol. 8, pp. 117593-117614, Jun. 2020.

\bibitem{ams}
J. Clancy, D. Mullins, B. Deegan, J. Horgan, E. Ward, C. Eising, P. Denny, E. Jones, M. Glavin,"Wireless Access for V2X Communications, Research, Challenges and Opportunities",\emph{ IEEE Communications Surveys and Tutorials}, vol. 26, no. 3, pp. 2082-2119, Apr. 2024.

\bibitem{ams}
A. Hajisami, J. Lansford, A. Dingankar and J. Misener, "A Tutorial on the LTE-V2X Direct Communication,"\emph{ IEEE Open Journal of Vehicular Technology}, vol. 3, pp. 388-398, Aug. 2022.

\bibitem{ams}
M. Muhammad and G.A. Safdar, "Survey on existing authentication issues for cellular-assisted V2X communication," \emph{Vehicular Communications}, vol. 12, pp. 50-65, Apr. 2018.

\bibitem{ams}
C. Shin, E. Farag, H. Ryu, M. Zhou and Y. Kim, "Vehicle-to-Everything (V2X) Evolution From 4G to 5G in 3GPP: Focusing on Resource Allocation Aspects," \emph{IEEE Access}, vol. 11, pp. 18689-18703, Feb. 2023.

\bibitem{fallgren2021book}
M. Fallgren, M. Dillinger, T. Mahmoodi, T. Svensson, and J. Wiley, Eds., \emph{Cellular V2X for Connected Automated Driving}. Hoboken, NJ, USA: Wiley, Apr. 2021.

\bibitem{rajalakshmi2024}
P. Rajalakshmi, "Towards 6G V2X Sidelink: Survey of Resource Allocation—Mathematical Formulations, Challenges, and Proposed Solutions," \emph{IEEE Open Journal of Vehicular Technology}, vol. 5, pp. 344–383, Feb. 2024.

\bibitem{ams}
H. Shakhatreh, A. H. Sawalmeh, A. Al-Fuqaha, Z. Dou, E. Almaita, I. Khalil, N. S. Othman, A. Khreishah, and M. Guizani, "Unmanned aerial vehicles (UAVs): A survey on civil applications and key research challenges," \emph{IEEE Access}, vol. 7, pp. 48572-48634, Apr. 2019.

\bibitem{ams}
L. Zhang, H. Zhao, S. Hou, Z. Zhao, H. Xu, X. Wu, Q. Wu, and R. Zhang, "A survey on 5G millimeter wave communications for UAV-assisted wireless networks,"\emph{IEEE Access}, vol. 7, pp. 117460-117504, Jul. 2019.

\bibitem{ams}
S. D. Muruganathan, X. Lin, H.-L. Maattanen, Z. Zou, W. A. Hapsari, and S. Yasukawa, "An overview of 3GPP release-15 study on enhanced LTE support for connected drones," \emph{IEEE Communications Standards Magazine}, vol. 5, no. 4, pp. 140-146, Dec. 2021.

\bibitem{ams}
L. Deng, G. Wu, J. Fu, Y. Zhang and Y. Yang, "Joint resource allocation and trajectory control for UAV-enabled vehicular communications," \emph{IEEE Access}, vol. 7, pp. 132806-132815, Sep. 2019.

\bibitem{ams}
R. Masroor, M. Naeem and W. Ejaz, "Resource management in UAV-assisted wireless networks: An optimization perspective," \textit{Ad Hoc Networks}, vol. 121, Art. no. 102596, Oct. 2021.

\bibitem{ams}
Y. Huang, M. Cui, G. Zhang and W. Chen, "Bandwidth, Power and Trajectory Optimization for UAV Base Station Networks With Backhaul and User QoS Constraints," \emph{IEEE Access}, vol. 8, pp. 67625-67634, Apr. 2020.

\bibitem{ams}
M. D. Nguyen, L. B. Le and A. Girard, "Integrated UAV Trajectory Control and Resource Allocation for UAV-Based Wireless Networks With Co-Channel Interference Management," \emph{IEEE Internet of Things Journal}, vol. 9, no. 14, pp. 12754-12769, Jul. 2022.

\bibitem{bhowmick2024}
A. Bhowmick, Y. K. Choukiker, I. Singh, and A. Nallanathan, Eds., \emph{5G and Beyond Wireless Communications: Fundamentals, Applications, and Challenges}. Boca Raton, FL, USA: CRC Press, 2024.

\bibitem{ams}
Z. Liu, G. Huang, Q. Zhong, H. Zheng and S. Zhao, "UAV-Aided Vehicular Communication Design With ‎Vehicle Trajectory’s Prediction," \emph{IEEE Wireless Communications Letters}, vol. 10, no. 6, pp. 1212-‎‎1216, Jun. 2021.‎

\bibitem{ams}
‎H. Dai, H. Bian, C. Li and B. Wang, "UAV-aided wireless ‎communication ‎design with energy constraint in space-air-ground ‎integrated green IoT ‎networks," \emph{IEEE Access}, vol. 8, pp. 86251-‎‎86261, May. 2020. ‎

\bibitem{ams}
‎M. Samir, M. Chraiti, C. Assi and A. Ghrayeb, "Joint Optimization of ‎UAV Trajectory and Radio Resource Allocation for Drive-Thru ‎Vehicular Networks," \emph{IEEE Wireless Communications and ‎Networking Conference (WCNC)}, Marrakesh, Morocco, pp. 1-6, Apr. 2019.‎

\bibitem{humayun2022}
M. Humayun, M. F. Almufareh, and N. Z. Jhanjhi, "Autonomous traffic system for emergency vehicles," \emph{Electronics}, vol. 11, no. 4, p. 510, Feb. 2022.

\bibitem{ams}
T. Wang, X. Pang, J. Tang, N. Zhao, X. Zhang, X. Wang, "Time and energy efficient data collection via UAV", \emph{Science China Information Sciences}, vol. 65, no. 8, Art. no. 182302, Jul. 2022.

\bibitem{ams}
Z. Ji, J. Tu, X. Guan, W. Yang, W. Yang, Q. Wu, "Energy Efficient Design in IRS-Assisted UAV Data Collection ‎System under Malicious Jamming", arXiv preprint arXiv:2208.14751, Aug. 2022.‎

\bibitem{ams}
Z. Zhang, G. Mao and B. D. O. Anderson, "Stochastic Characterization of Information Propagation Process in Vehicular Ad hoc Networks,"\emph{ IEEE Transactions on Intelligent Transportation Systems}, vol. 15, no. 1, pp. 122-135, Feb. 2014.

\bibitem{ams}
M. Samir, S. Sharafeddine, C. Assi, T. M. Nguyen and A. Ghrayeb, "Trajectory Planning and Resource Allocation of Multiple UAVs for Data Delivery in Vehicular Networks," \emph{IEEE Networking Letters}, vol. 1, ‎no. 3, pp. 107-110, Sept. 2019.

\bibitem{ams}
Y. Zeng, J. Xu and R. Zhang, "Energy Minimization for Wireless Communication With Rotary-Wing UAV," \emph{IEEE Transactions on Wireless Communications}, vol. 18, no. 4, pp. 2329-2345, Apr. 2019.

\bibitem{boyd2004}
S. Boyd and L. Vandenberghe, \emph{Convex Optimization}. Cambridge, UK: Cambridge University Press, Mar. 2004.

\bibitem{ams}
G. Zhang, Q. Wu, M. Cui and R. Zhang, "Securing UAV Communications via Joint Trajectory and Power Control," \emph{IEEE Transactions on Wireless Communications}, vol. 18, no. 2, pp. 1376-1389, Feb. 2019.

\bibitem{ams}
Y. Chen, Y. Wang, J. Zhang and M. D. Renzo, "QoS-Driven Spectrum Sharing for Reconfigurable Intelligent Surfaces (RISs) Aided Vehicular Networks," \emph{IEEE Transactions on Wireless Communications}, vol. 20, no. 9, pp. 5969-5985, Sept. 2021.

\bibitem{ams}
X. Jiang, Z. Yang, N. Zhao, Y. Chen, Z. Ding and X. Wang, "Resource ‎Allocation and Trajectory Optimization for UAV-Enabled Multi-User ‎Covert Communications," \emph{IEEE Transactions on Vehicular ‎Technology}, vol. 70, no. 2, pp. 1989-1994, Feb. 2021.‎

\bibitem{sun2024}
G. Sun, Y. Wang, Z. Sun, L. He, and X. Zheng, "Joint Task Offloading and Trajectory Control for Multi-UAV-Assisted Mobile Edge Computing," in \emph{Proc. IEEE International Conference on Communications (ICC)}, Denver, CO, USA, pp. 2652--2657, 2024.

\bibitem{ams}
R. Zhang, R. Lu, X. Cheng, N. Wang and L. Yang, "A UAV-Enabled Data Dissemination Protocol With Proactive Caching and File Sharing in V2X Networks," \emph{IEEE Transactions on Communications}, vol. 69, no. 6, pp. 3930-3942, June 2021.

\bibitem{ams}
G. Yang, R. Dai and Y. -C. Liang, "Energy-Efficient UAV Backscatter Communication With Joint Trajectory Design and Resource Optimization," \emph{IEEE Transactions on Wireless Communications}, vol. 20, no. 2, pp. 926-941, Feb. 2021.

\end{thebibliography}
\end{document}